# Probing the structure, morphology and multifold blue absorption of a new red-emitting nanophosphor for LEDs

**Savvi Mishra,[1] Isha Bharti,[1] N. Vijayan,[1] R. K. Sharma,[1] L. M. Kandpal,[1] V. Shanker,[1] M. K. Dalai,[1] R. Rajeswari,[2] C. K. Jayasankar,[2] S. Surendra Babu[3] and D. Haranath,[1,a)]**

[1]*CSIR-National Physical Laboratory, Dr K S Krishnan Road, New Delhi – 110 012, India.*
[2]*Department of Physics, Sri Venkateswara University, Tirupati-517502, India.*
[3]*Directorate of Laser Systems, Research Centre IMARAT, Vignyana Kancha, Hyderabad-500069, India.*

**Abstract**

There has been a stringent demand for blue (~450-470 nm) absorbing and red (~611 nm) emitting material system in phosphor converted white light emitting diodes (WLEDs) available in the market. Conventionally used red-emitting $Y_2O_3$:$Eu^{3+}$phosphor has negligible absorption for blue light produced by GaInN based LED chip. To address this issue, a new red-emitting $Gd_2CaZnO_5$:$Eu^{3+}$ (GCZO:$Eu^{3+}$) nanophosphor system having exceptionally strong absorption for blue (~465 nm) and significant red (~611 nm) photoluminescence (PL) is presented. This is attributed to a dominant f-f transition ($^5D_0 \rightarrow ^7F_2$) of $Eu^{3+}$ ions, aroused due to an efficient energy transfer from the $Gd^{3+}$ sites of the host lattice to $Eu^{3+}$ ions. X-ray diffraction and microscopy observations revealed the nanocrystalline nature and a bit elongated morphology of the sample respectively. While the energy dispersive x-ray analysis identified the chemical constituents of the GCZO:$Eu^{3+}$ nanophosphor, the color overlay imaging confirmed the substitution of $Eu^{3+}$ for $Gd^{3+}$ ions. It is highly anticipated that the multifold absorption at ~465 nm would certainly improve the color rendering properties of existing WLEDs.

**Introduction**

White light emitting diodes (WLEDs) are the future of solid-state lighting because of their innumerable advantages starting from being energy efficient, economical, compact, durable and having high luminous efficiency as well. The most widely used WLED uses a yellow-emitting $Y_3Al_5O_{12}$:$Ce^{3+}$ phosphor integrated to a blue-emitting GaInN diode, a combination of which produces white light. However, some problems creep in regarding the color output of the WLEDs having very high color temperature (>10000 K) and low color rendering index (<60). The reason being the red-emitting $Y_2O_3$:$Eu^{3+}$ phosphor employed in WLEDs has a negligible absorption in the blue (450-470 nm) region. Hence, there is a need of a newer red-emitting phosphor material with strong absorption under blue light to render better color temperature and thereafter an ideal WLED will be realized.

Inorganic phosphors have been broadly investigated in various aspects of lighting applications such as screens in field emission displays, plasma display panel thin film electroluminescent devices and WLEDs and displays to a name a few [1-5]. Being chemically stable, moisture resistant etc. oxide phosphors have overpowered sulphide phosphors [6-7] in these phosphor converted WLEDs. Out of which rare-earth ions doped oxide lattices have been worked upon extensively as they exhibit excellent luminescent properties and their f-f line transitions in visible range result in high efficiency [8].

To the best of our knowledge, not much has been reported on rare-earth doped $A_2BMO_5$ type ternary oxides as efficient phosphors in the literature. Michel *et al.* [9] worked on $Ln_{4-2x}Ba_{2+2x}Zn_{2-x}O_{10-2x}$ (where, Ln=La, Nd, and *x*=0-0.25) and studied their crystal structures. Michel and Raveau [9-11] studied the structural characterization of $A_2BaCuO_5$ and $A_2BaZnO_5$ systems (where, A=rare-earth and M=Cu, Zn). $A_2BMO_5$ type oxide systems demonstrate unusual yet interesting structural, physical and chemical properties along with unique optical, magnetic and superconducting properties [12-13]. Europium is one of the most popular dopant in luminescent materials used in various lighting applications [14]. Recently, Lammers *et al.* [15] have reported red-emitting $Eu^{3+}$ doped $Gd_2BaZnO_5$ with 75% external quantum efficiency.

Earlier studies exemplify that $A_2BMO_5$ systems were synthesized by conventional solid-state reaction method at temperatures >1200ºC, but for the first time we have employed a self-proliferating sol-gel combustion technique to prepare the phosphor in nano regime with a yield >90%, which is far better than earlier reports. Further, the achievement of a very narrow size distribution ($\sigma$=0.2-0.5) of nanophosphor particles is another key feature of the presented work.

In this letter we report a new promising GCZO:$Eu^{3+}$ nanophosphor with a strong red emission excitable by blue light from GaInN LED. Further this nanophosphor has high density and large band gap similar to its isostructural compound $BaY_2ZnO_5$:$Eu^{3+}$ [16,17].

**Experimental procedure**

**Preparation of GCZO:Eu$^{3+}$ nanophosphor:** GCZO:Eu$^{3+}$ samples were prepared using a self-proliferating sol-gel combustion method. The flow chart of the nanophosphor synthesis is shown in figure 1. Being a self-propagating reaction, this synthesis technique utilizes the exothermicity of the redox reactions involved and a very narrow size distribution of nanophosphor is conceived. Among various solution routes for oxide nanomaterial synthesis, this method is versatile in producing uniform and ultrafine particles as end product in negligible span (~15 min) of firing times [18-19]. In a typical experiment, clear nitrate solutions of all the precursor salts were chelated with citric acid solution and a non-luminescent gel under ultra-violet (254 nm) is formed. The dual role of citric acid is being a fuel and a gelling agent. Finally the gel is introduced in a preheated furnace at ~800ºC for 15 minutes in flowing oxygen atmosphere to obtain the stand-alone GCZO:Eu$^{3+}$ nanophosphor.

**Characterization:** The phase of the nanophosphor was identified by x-ray powder diffractometer of Bruker D-8 make with Cu Kα radiation operated at 35 kV and 30 mA. The photoluminescence (PL) spectra were recorded at room temperature using an Edinburgh Luminescence Spectrometer (Model: F900) equipped with a xenon lamp in the scan range from 200-800 nm. To investigate thoroughly about the morphology and particle size of the synthesized nanophosphors, High Resolution transmission electron microscope (HRTEM, model number TECHNAI G20-Twin, 200 kV) was used which was equipped with energy dispersive X-ray analysis (EDAX) attachment as well. The color overlay imaging was performed using a reflection based high-resolution time of flight secondary ion mass spectroscopy (ToF-SIMS) of ION TOF GmbH, Germany. ToF-

SIMS is an advanced technique useful for identification of chemical constituents' when they are incorporated in low concentrations.

**Figure 1.** (Color online) Flow chart depicting the synthesis of GCZO:Eu$^{3+}$ nanophosphor.

## Results and Discussion

### Crystal structure of the sample

Figure 1 shows the x-ray diffraction (XRD) pattern of GCZO:Eu$^{3+}$ nanophosphor along with the particular (h k l) planes at room temperature with X-ray wavelength of 1.54060 Å. The sharp crystalline features observed in the XRD profiles are due to the high reaction temperature provided for a short interval of time for the nanophosphor samples during the combustion process. Since the nanomaterial under study, Gd$_2$CaZnO$_5$ (GCZO) is an entirely new lattice and the respective JCPDS data is not available, we have used WIN-INDEX (ver. 3.08) software for the structure refinement studies and determination of (h k l) values corresponding to the crystalline planes. The A$_2$BMO$_5$ system is found in three structural types which are discussed elsewhere [20]. It is proposed that the new lattice GCZO belongs to type II structure, more precisely orthorhombic structure with space group *Pbnm* [21].

Michel *et al.* [9] worked on Ln$_{4-2x}$Ba$_{2+2x}$Zn$_{2-x}$O$_{10-2x}$ (where, Ln=La, Nd, and *x*=0-0.25) system and studied their crystal structures similar to A$_2$BaZnO$_5$ systems (where, A=rare-earth and M=Cu, Zn). They could not distinguish between Ba$^{2+}$ and Ln$^{3+}$ ions, and proposed Ba$^{2+}$ ions occupy sites with larger volume. In their final compound layers of Ln$_2$O$_5$ were present that formed face and edge sharing LnO$_8$ polyhedra, and Ba$^{2+}$ and Zn$^{2+}$ were inserted in between. Lanthanide ion's site symmetry is C$_{2v}$ whereas in the current case Gd$^{3+}$ has two preferential crystallographic sites with low (C$_s$) and high (C$_{2v}$) crystal field symmetries. From this it could be deduced that the structure consists of $GdO_7$, $CaO_{11}$, and $ZnO_5$ polyhedra. Each Gd atom, which is coordinated seven fold by oxygen atoms, forms a trigonal prism and two such prisms should be present with different Gd-O distances [9, 22-23].

Considering the ionic radii and valence states of $Gd^{3+}$ and $Eu^{3+}$ ions, it could be presumed that $Gd^{3+}$ ions are being substituted by $Eu^{3+}$ ions in the GCZO lattice giving strong red emission at ~611 nm. However, there is always a chance of substitution of Eu3+ ions for Ca and Zn sites also. This possibility has been paid attention and discussed in the latter sections. Further it was confirmed that $Eu^{3+}$ forms two centers as it substitutes for two crystallographic $Gd^{3+}$ sites.

**Figure 2. (**Color online) XRD profile of GCZO:$Eu^{3+}$ nanophosphor at room temperature. The (h k l) values generated from WIN-INDEX (ver. 3.08) software are shown in the figure.

**Electron microscopy and EDAX analysis**

The as-synthesized nanophosphor samples were subjected to transmission electron microscopy (TEM) observations at low (80 kX) and high (200 kX) magnifications. Typical TEM and high-resolution TEM (HRTEM) micrographs of the nanophosphor samples are shown in figure 3(a-d). Fluffy, voluminous and delicate porous nanophosphor samples are formed due to evolution of large amount of gases during the self-proliferating sol-gel combustion reaction. Loosely bound agglomerates of nanophosphor particles are formed that have uniform size distributions. At low (80 kX) magnification, long chain-like morphology with polygon shaped nanostructures and sharp boundaries are evidently seen in figure 3 (a-b). At higher (200 kX) magnification, the fringes are clearly apparent that reflect the strong crystalline nature of the nanophosphor particles. Chain-like nanostructures may be attributed to the high magnetic moment of the rare-earth ($Gd^{3+}$) ions involved. From TEM observations the average particle size has been attributed to be in the range 10-20 nm.

**Figure 3(a-d).** (Color online) TEM and HRTEM micrographs taken at various magnifications with their corresponding scales.

Detailed elemental analysis of the nanophosphor was corroborated by energy dispersive x-ray analysis (EDAX) shown in Figure 4. Spot EDAX measurement is done with reduced beam size and low accelerating potential so as to have increased signal to noise ratio. EDAX spectrum revealed the presence of all the lattice elements with dopants as well. All the significant peaks of Gadolinium, Calcium, Zinc, and Europium are present in the spectrum and are represented in the figure accordingly.

**Figure 4.** (Color online) EDAX spectrum depicting the composition of various elements present in the GCZO:$Eu^{3+}$ nanophosphor.

**Color overlay analysis**

As it was mentioned earlier that the $Eu^{3+}$ ions could have a chance of substituting for either Gd, Ca or Zn in the GCZO:$Eu^{3+}$ nanophosphor system. If we consider the ionic radii and valence states of these ions, the most preferable site is that of $Gd^{3+}$. To confirm this we have carried out the color overlay analysis of ToF-SIMS technique. For this a thick film of GCZO:$Eu^{3+}$ nanophosphor is made using an organic binder comprising of nitrocellulose and amyl acetate over an ITO-coated conducting glass substrate. The spectral analyses of the nanophosphor in both positive and negative ion detection modes were investigated. Chemical imaging mode of the ToF-SIMS instrument was used to verify the homogenous doping of $Eu^{3+}$ ions for $Gd^{3+}$ sites in the host matrix. Arbitrary red and green colors were assigned to Eu and Gd elements respectively, and color overlay was performed as shown in Figure 5. As both elements complement to each other very-well, the color of the final product was observed to be the secondary (mixed) color. The degree of presence of mixed color signifies the homogeneity of the dopant distribution in the nanophosphor sample. However, the case was not similar with other precursor elements such as Zn and Ca (not shown) present in the $GCZO:Eu^{3+}$ nanophosphor samples.

**Figure 5. (**Color online) Correlation analysis using two color overlay images confirming the substitution of Eu for Gd sites.

**Photoluminescence (PL) studies of the nanophosphor**

The room-temperature photoluminescence excitation (PLE) spectrum of $GCZO:Eu^{3+}$ nanophosphor recorded in the wavelength range of 200-500 nm with emission monitored at ~611 nm is shown in Figure 6. It is clearly seen from the figure that the PLE spectrum consists of overlapped charge transfer (CT) band (whose maximum is at ~250 nm) and $Gd^{3+}-O^{2-}$ transitions. CT corresponding to $Eu^{3+}-O^{2-}$ and also to $^8S_{7/2} \rightarrow {}^6I_J$, $^8S_{7/2} \rightarrow {}^6D_J$ and $^8S_{7/2} \rightarrow {}^6P_J$ (J=1, 2 and 3) transitions analogous to $Gd^{3+}$ transitions. Sharp well-known major line transitions due to f-f transitions of $Eu^{3+}$ and $Gd^{3+}$ are quiet visible in the PL spectra. $Gd^{3+}$ transitions are mostly in the overlapped region but $Eu^{3+}$ can be seen in the range 350-470 nm range. It is interesting to note that the excitation peak intensities corresponding to ~395 and ~465 nm are quite prominent in the GCZO nanophosphor system, which are not usually found in commercial $Y_2O_3:Eu^{3+}$ and $Y_2O_2S:Eu^{3+}$ bulk-phosphors that are currently being used as red-emitting phosphor components in WLEDs. These excitation lines are due to $^7F_0 \rightarrow ^5L_6$ and $^7F_0 \rightarrow ^5D_2$ transitions of $Eu^{3+}$ levels in the near UV (~395 nm) and blue (~465 nm) regions, respectively. The commercial red-emitting phosphors mentioned above have serious drawback of negligible absorption for blue (~450-470 nm) radiations emitted by GaInN blue-LED chip. Hence, from the above results, we suggest that the GCZO:$Eu^{3+}$ nanophosphor system is a unique substitute to the existing bulk phosphors in improving the color rendition of WLEDs as they have multifold absorption at ~465 nm. The dotted line depicted in figure 6 highlights the multifold absorption at ~465 nm. Table 1 shows the comparative datasheet of commercial $Y_2O_3$:Eu (bulk), GCZO:Eu (bulk and nano) phosphors, ratio of 465 nm absorption to CT band and their respective percentage of absorption. From the PLE spectrum of the GCZO:$Eu^{3+}$ nanophosphor, an efficient energy transfer between $Eu^{3+}$ and $O^{2-}$ ions; and between $Gd^{3+}$ and $Eu^{3+}$ ions that is predicted in the UV (~248 nm) region can also be noticed. The room-temperature steady-state PL spectrum was recorded in the wavelength range 500-700 nm under the excitation wavelength of 248 nm. It is distinctly seen that the PL spectrum consists of several sharp lines corresponding to the transitions between $^5D_0$ and $^7F_J$ (J=1, 2 and 3) levels. The lines due to $^5D_0 \rightarrow ^7F_2$ are most dominating in the spectrum corresponding to ~611 nm although other peaks due to $^5D_0 \rightarrow ^7F_1$ and $^5D_0 \rightarrow ^7F_3$ are comparatively weak [33]. As can be seen from the PL spectrum it is inferred that intensities of $^5D_0$ emission lines of $Eu^{3+}$ overpower $^5D_J$ emission lines at room temperature according to temperature dependence for multiphonon relaxation [25] and also when it is excited by the wavelength corresponding to charge transfer band [26]. It is also anticipated that as $Gd^{3+}$- $^6I$ levels overlap completely with $Eu^{3+}$-$O^{2-}$ charge transfer band, $Eu^{3+}$ is effectively trapping energy from $Gd^{3+}$-$^6I$ transitions [27-28].

**Figure 6. (**Color online) (a) Photoluminescence excitation (PLE) of GCZO:$Eu^{3+}$ nanophosphor monitored at ~611 nm emission and (b) PL spectra recorded at ~248 nm. The dotted line highlights the multifold absorption at ~465 nm.

## Conclusions

In summary, a simple and viable method of making high (>90%) yield GCZO:$Eu^{3+}$ nanophosphor with exceptional blue (~465 nm) absorption is presented. Strong absorption in the blue region is attributed to the quantum size effects. The control over morphology and uniform size distribution (10-20 nm) of nanophosphor particles has been achieved quantitatively. XRD, TEM, EDAX and color overlay analysis were performed to verify the crystalline phase, particle size, elemental composition and dopant distribution, respectively. Finally, the GCZO:$Eu^{3+}$ nanophosphor may find a plausible replacement for currently used and weakly absorbing phosphors in WLEDs thus producing ideal white-light.

## Abbreviations

XRD, x-ray diffraction; WLED, white light emitting diodes; PL, photoluminescence;

## Competing interests

Authors have no competing interests.

**Authors' contributions**

The authors SM and IB prepared the nanophosphor samples; RKS and LMK are involved in the preparation and characterization of bulk samples; NV analysed the XRD data; MKD performed the color overlay analysis; VS and DH had performed the PL studies and drafted the manuscript; RR, CKJ and SSB have contributed immensely for discussions while preparing the manuscript. Finally all the authors' have read and approved the manuscript.

**Acknowledgments**

The authors (DH and SM) gratefully acknowledge the Department of Science and Technology (DST), Government of India for providing financial assistance under the scheme # SR/FTP/PS-012/2010 to carryout the research work.

## Figure Captions

**Figure 1.** (Color online) Flow chart depicting the synthesis of GCZO:Eu$^{3+}$ nanophosphor.

**Figure 2.** (Color online) XRD profile of GCZO:Eu$^{3+}$ nanophosphor at room temperature. The (h k l) values generated from WIN-INDEX (ver. 3.08) software are shown in the figure.

**Figure 3(a-d).** (Color online) TEM and HRTEM micrographs taken at various magnifications with their corresponding scales.

**Figure 4.** (Color online) EDAX spectrum depicting the composition of various elements present in the GCZO:Eu$^{3+}$ nanophosphor.

**Figure 5.** (Color online) Correlation analysis using two color overlay images confirming the substitution of Eu for Gd sites.

**Figure 6.** (Color online) (a) Photoluminescence excitation (PLE) of GCZO:Eu$^{3+}$ nanophosphor monitored at ~611 nm emission and (b) PL spectra recorded at ~248 nm. The dotted line highlights the multifold absorption at ~465 nm.

**Table 1** shows the comparative datasheet of commercial $Y_2O_3$:Eu (bulk), GCZO:Eu (bulk and nano) phosphors, ratio of 465 nm absorption to CT band and their respective percentage of absorption.

| Name of the phosphor | Intensity ratio of 395 nm transition and CTB | Intensity ratio of 465 nm transition and CTB | Relative absorption % at 465 nm |
|---|---|---|---|
| $Y_2O_3$:Eu (bulk) | 0.127/3.62 | 0.108/3.62 | 2.98% |
| GCZO:Eu (bulk) | 0.117/3.56 | 0.211/3.56 | 5.92% |
| GCZO:Eu (nano) | 1.39/1.16 | 0.727/1.16 | 62.6% |

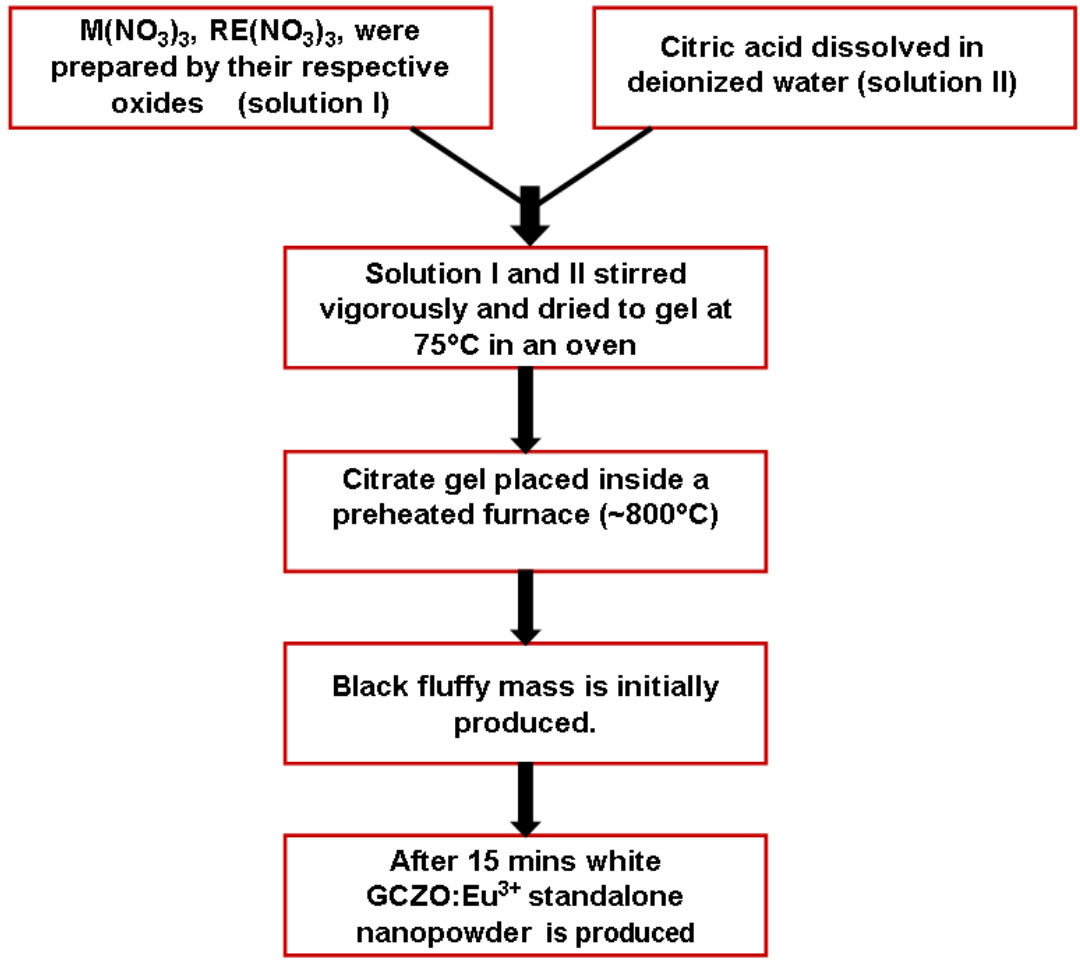

**Figure 1**

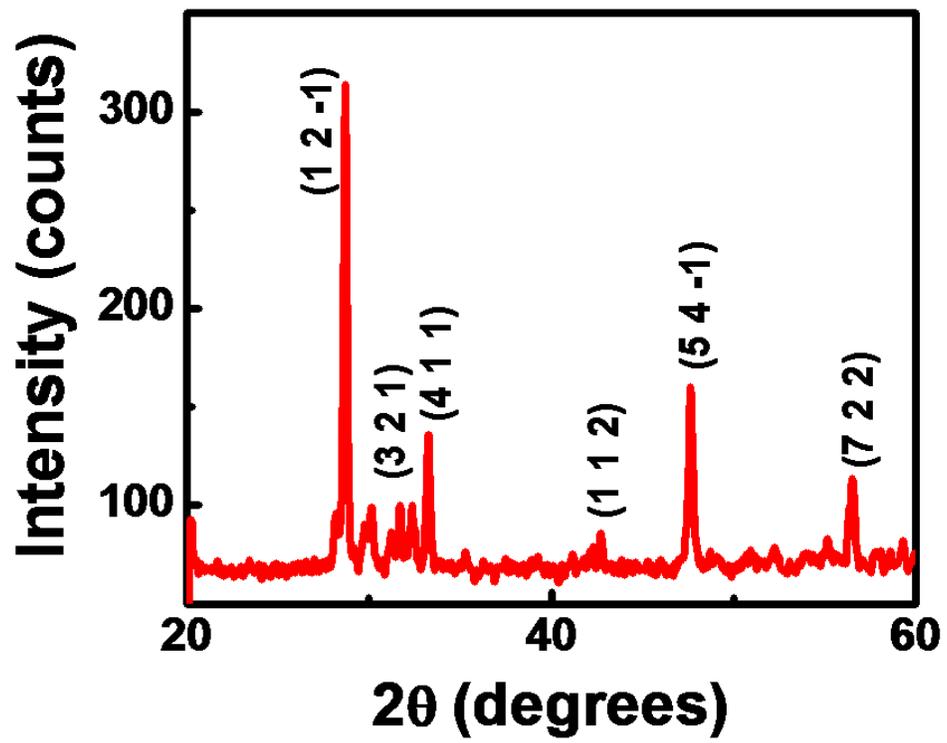

**Figure 2**

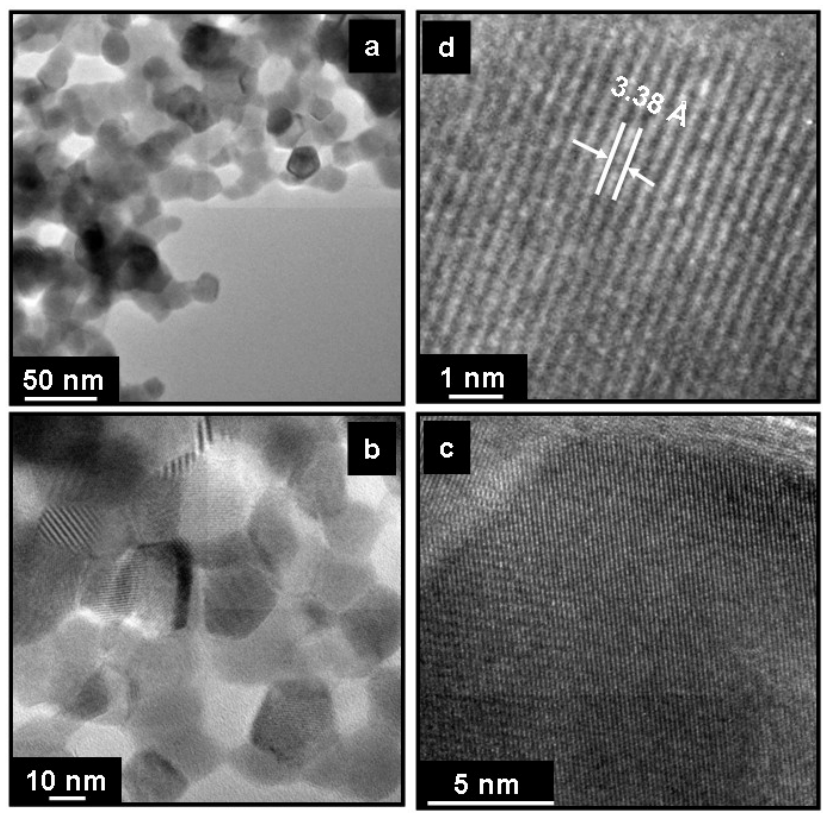

**Figure 3**

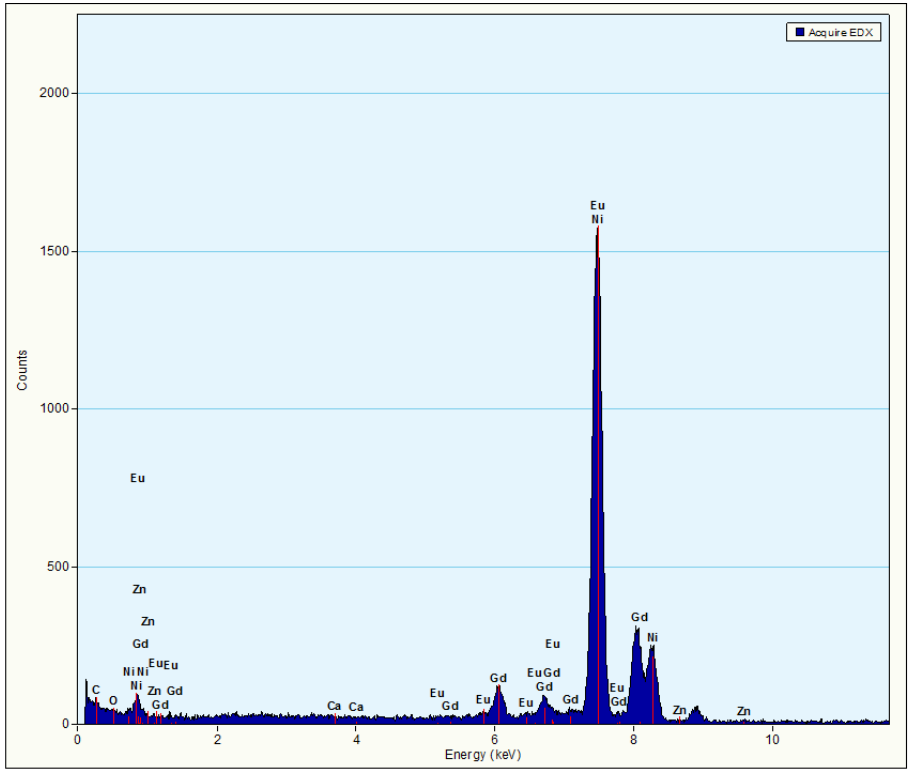

**Figure 4**

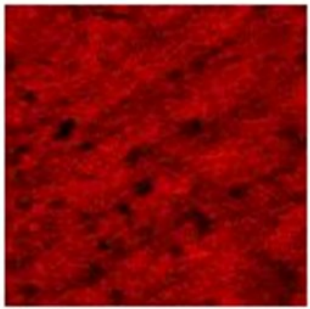

Eu+
mc:101 tc:8.21e+5

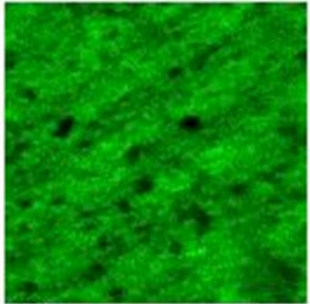

Gd+
mc:142 tc:1.15e+6

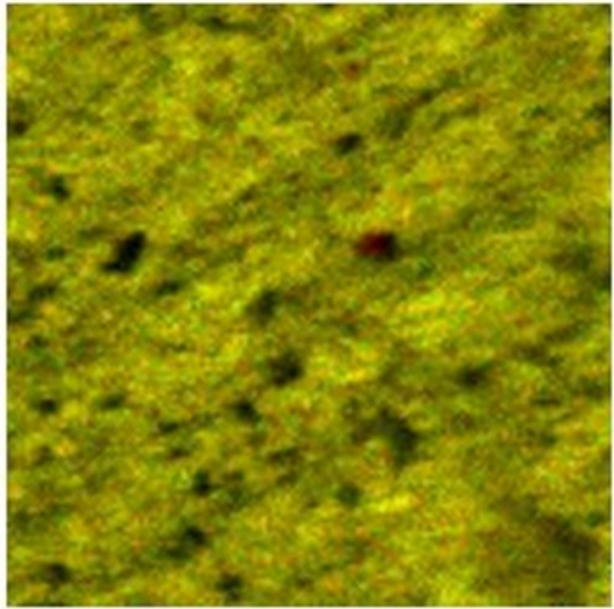

Overlay: Eu+ , Gd+ ,

**Figure 5**

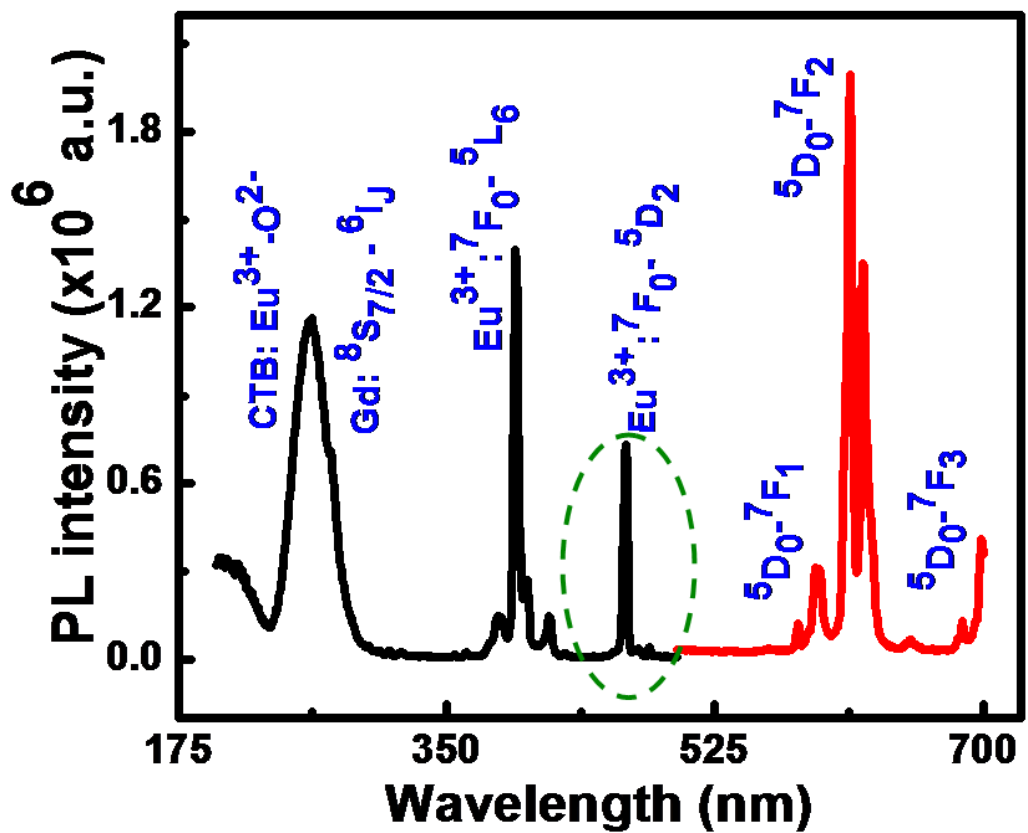

**Figure 6**